\documentstyle[aps,twocolumn,epsf]{revtex}

\begin{document}


\preprint{UGI-98-25}

\title{Resummation of soft modes in the free energy of $\Phi^4$ 
theory\thanks{Talk given at the 5th International Workshop on Thermal Field 
Theories and Their Applications, Regensburg, Germany, 10.-14.~August 1998.}}

\author{Stefan Leupold}

\address{Institut f\"ur Theoretische Physik, Universit\"at Giessen,\\
D-35392 Giessen, Germany}

\date{August 26, 1998}

\maketitle

\begin{abstract}
A new method is proposed for the calculation of the free 
energy of an $N$-component $\Phi^4$ theory at finite temperature.
The method combines a perturbative treatment of the hard 
modes with a non-perturbative treatment in the effectively 
three-dimensional sector of the soft modes. The separation 
between hard and soft modes is achieved by the effective
field theory method of Braaten and Nieto \cite{BN95}. 
One- and two-loop gap equations for the screening mass are
used to resum higher order effects in the soft sector. 
The proposed method is similar to the screened perturbation
theory of Karsch, Patk\'os, and Petreczky 
\cite{Ka97} albeit avoiding the difficulty 
of solving gap equations in the full four-dimensional theory.
Since gap equations are only used in the three-dimensional
effective theory the tedious evaluation of Matsubara sums 
is not necessary. This simplification makes it possible to
go beyond the one-loop gap equation. 
The large $N$ as well as the finite $N$ case are discussed. 
\end{abstract}


\narrowtext
\vspace{1cm}

\begin{figure}
\epsfxsize= 8.8cm
\centerline{\epsfbox{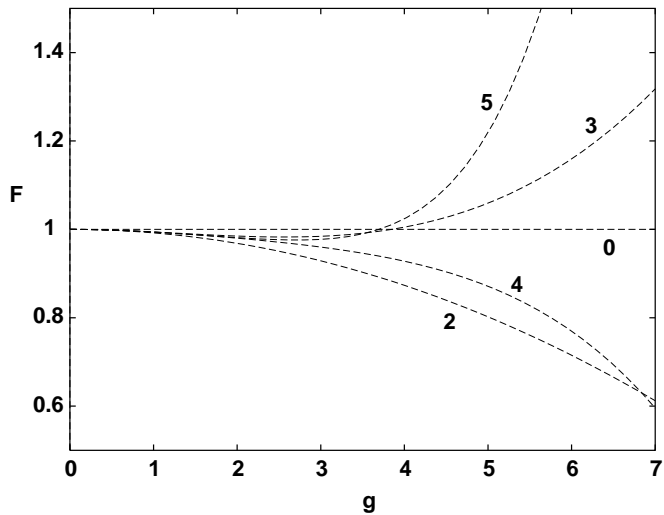}}
\vskip4mm
\caption{The perturbative result for the free energy density
of a one-component massless
$g^2 \Phi^4$ theory as a function of the coupling $g$. The label $n$ marks the
highest perturbative contribution $\sim g^n$ which is taken into account. 
The free energy density is normalized to its ideal gas value.}
\label{fig:pert}
\end{figure}

In the last few years there has been a lot of progress in the 
calculation of the free energy of high temperature field theories 
\cite{BN95,Ka97,CP94,Pa94,PC94,AZ94,AZ95,ZK95,BN96,AHLW97,FST92,PS95,KLPR97,%
Kast97,Ha97,DHLR96,RS97,DHLR97,PKPS98,VB98,BLNR98,Re98,Sc98}. 
On the perturbative level all corrections to the ideal gas up to order $g^5$
have been calculated for gauge theories \cite{CP94,Pa94,PC94,AZ94,AZ95,ZK95,BN96} as
well as for scalar $\Phi^4$ theory \cite{FST92,PS95,BN95}. Unfortunately,
these calculations revealed that already for moderate values of the 
coupling constant higher
order corrections become large and oscillating, i.e.~each additional higher order
correction flips the sign of the deviation from the ideal gas limit 
(cf.~fig.~\ref{fig:pert}). In addition,
the results depend crucially on the chosen renormalization
point \cite{ZK95,BN96}. 

In this talk I would like to contribute to the question whether/how the perturbative
calculation of the free energy can be improved such that one gets a reliable answer
not only for very small values of the coupling constant (where perturbation theory
works, but there one might use the free (unperturbed) result as well), but also for 
moderate values of $g$. The aim is to reorganize the loop expansion such that
higher loop corrections do not drastically change the result. Of course, it is not
clear whether such a reorganization is possible or if one has to rely solely on
genuine non-perturbative approaches like lattice calculations
\cite{Creutz}. Even for theories
where lattice calculations yield reliable results an improved perturbative 
calculation scheme is nonetheless useful to identify the relevant degrees of 
freedom of the system. Indeed, the analysis of QCD lattice calculations indicates
that at high temperatures the system of {\it massless coupled} modes can be very 
well approximated by a gas of {\it noninteracting massive} particles 
\cite{Ka89,GS93,PK96,LH97}. It is clearly desirable to support this heuristic
finding by a calculation starting from first principles such that also the fit
parameters (e.g.~the temperature dependence of the mass) can be calculated from
the underlying theory. In this spirit a reorganization of the loop expansion
called {\it screened perturbation theory}
was suggested in \cite{Ka97} for an $N$-component scalar $\Phi^4$ theory. 

\begin{figure}
\epsfxsize= 8.8cm
\centerline{\epsfbox{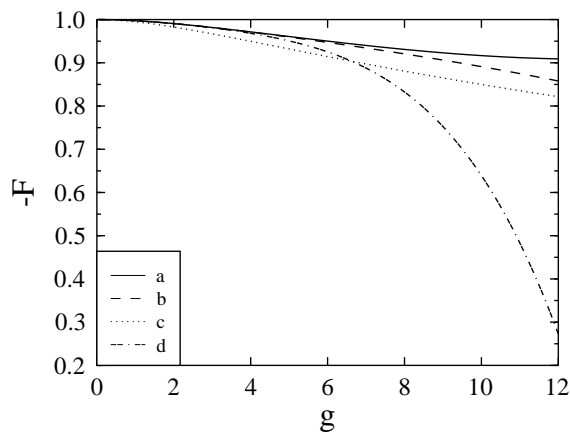}}
\vskip4mm
\caption{The free energy density of the scalar field theory in the large $N$ limit
as a function of the coupling constant in units of the free energy density 
of a massless 
ideal gas. The curves represent three-loop (a), two-loop (b) and one-loop (c) 
results of 
the screened loop expansion as well as the $O(g^4)$ result (d) of the conventional
perturbative expansion. For details see \protect\cite{Ka97} where the figure is 
taken from.}
\label{fig:karsch}
\end{figure}

Starting from the Lagrangian
\begin{equation}
\label{eq:orglagr}
{\cal L} = {1\over 2} (\partial \Phi_i)^2 + {g^2 \over 24 N} (\Phi_i^2)^2  
\end{equation}
(where $i$ labels the $N$ components of the fields) reorganization of perturbation
theory is achieved by adding and subtracting a mass term:
\begin{eqnarray}
\label{eq:karschlagr}
{\cal L} &=& {\cal L}_0 + {\cal L}_{\rm int}  \,, \nonumber \\
{\cal L}_0 &=& {1\over 2} (\partial \Phi_i)^2 + {1 \over 2} M^2 \Phi_i^2  \,, \\
{\cal L}_{\rm int} &=& 
- {1 \over 2} M^2 \Phi_i^2 + {g^2 \over 24 N} (\Phi_i^2)^2 \,. 
\nonumber
\end{eqnarray}
The mass $M$ is calculated from the one-loop gap equation
\begin{equation}
  \label{eq:olgkarsch}
M^2 = 
\begin{minipage}[b]{3em}
\begin{figure}
\epsfxsize= 2em
\centerline{\epsfbox{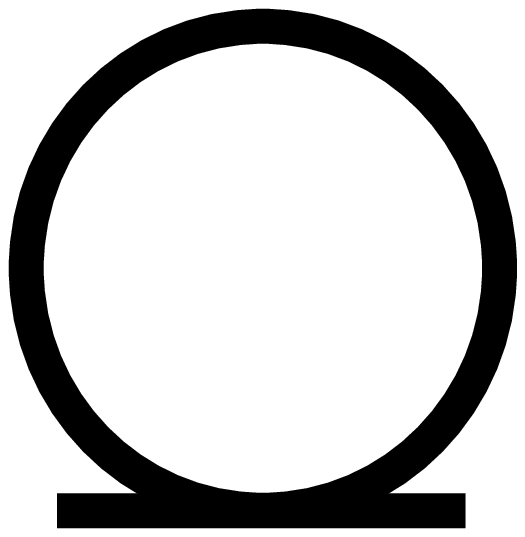}}
\end{figure}
\end{minipage}   \;.
\end{equation}
In \cite{Ka97} the necessary loop calculations were performed up to three loops
for the large $N$ case and up to two loops for arbitrary $N$. The result for the
free energy density for the large $N$ case is shown in 
fig.~\ref{fig:karsch} taken from
\cite{Ka97}. Obviously the convergence of the screened perturbation theory looks
very promising even for large values of the coupling constant. 

\begin{figure}
\epsfxsize= 3cm
\centerline{\epsfbox{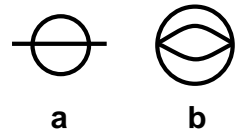}}
\caption{Sunset (a) and basketball (b) diagram contributing to the gap equation for
the mass and to the free energy density, respectively.}
\label{fig:basket}
\end{figure}

Of course, it is 
interesting to check whether this scheme works equally well for arbitrary $N$. 
There is a
technical reason why the three-loop calculations were not performed in \cite{Ka97}
for small $N$: It is very complicated to calculate the sunset and the 
basketball 
diagram shown in fig.~\ref{fig:basket} a and b, respectively. While the latter
contributes to the free energy, the former shows up as a two-loop correction to
the gap equation (\ref{eq:olgkarsch}). In the large $N$ limit both diagrams are
suppressed by a factor $1/N$. For finite $N$, however, these diagrams have to be 
taken into account. Unfortunately, to calculate these diagrams with massive
propagators is quite messy especially to perform the Matsubara sums. 

To circumvent this technical problem the following observation from
the QCD case might be useful: 
If the (oscillating) perturbative corrections to the massless ideal gas
are separated in their contributions from the hard (order of temperature $T$) 
and from the soft (order $gT$) sector \cite{BN96} one observes that the 
mentioned oscillatory
behavior is mainly caused by the soft modes, i.e.~for not too large values of the
coupling constant the contribution of the hard modes to the free energy is well
behaved while the one from the soft modes is not. The latter, however, can be 
described by an effective three-dimensional (vacuum like) theory. 
Hence, if non-perturbative
improvements are restricted to the soft mode sector, Matsubara sums do not show
up and one can use the techniques of vacuum perturbation theory to calculate the
necessary integrals which enter the respective gap equation and the free energy
calculation. This consideration
is the basis of the work presented here. 

The recipe to calculate the free energy density is the following:\\
1. Disentangle the contributions from different scales (hard and soft). \\
2. Use conventional perturbation theory to calculate the contribution from the hard
scale. \\
3. Use screened perturbation theory to calculate the contribution from the soft
scale. 

Fortunately, the first two points have already been performed for one-component
$\Phi^4$ theory in \cite{BN95} and for gauge theories in \cite{FKRS94}. For
simplicity and to compare
my results with the four-dimensional screened perturbation theory developed in
\cite{Ka97} I will work in the following with an $N$-component $\Phi^4$ theory 
which is a straightforward generalization of the system studied in \cite{BN95}. 
The free energy density receives contributions from the hard modes, $F_h$, as well
as from the soft modes, $F_s$. In conventional perturbation theory up to $O(g^5)$
these contributions are schematically given by\footnote{Strictly speaking the 
number sign in front of $g^5$ represents not a pure number but 
contains also a log$\,g$ contribution.} 
\begin{eqnarray}
  \label{eq:fepertsch}
F_h &=& \# + \# g^2 \phantom{+\# g^3} + \# g^4 \phantom{+ \# g^5}  \\
F_s &=& \phantom{\# + \# g^2 +} \# g^3 + \# g^4 + \# g^5  
\label{eq:fepertsoft}
\end{eqnarray}
The perturbative contribution from the hard sector is given by (cf.~\cite{BN95})
\begin{equation}
  \label{eq:hardtot}
F_h = F_h^{1l} + F_h^{2l} + F_h^{3l}
\end{equation}
with
\begin{equation}
  \label{eq:idealh}
F_h^{1l} = -{N\pi^2 T^4 \over 90} \,,
\end{equation}
\begin{equation}
  \label{eq:2lh}
F_h^{2l} = {1 \over 8} g^2(\mu) \tilde N {T^4 \over 12^2}  \,,
\end{equation}
\begin{eqnarray}
\label{eq:Fh3l}
\lefteqn{F_h^{3l} = 
-{1\over 72} \pi^2 T^4 {\tilde N \over N} \left({g^2\over 16\pi^2}\right)^2}
 \\ &\times & 
\left[ 
{N+8 \over 3} \log{\mu\over4\pi T} + \tilde N \gamma + {31\over 15} 
+ 4{\xi'(-1) \over \xi(-1)} -2{\xi'(-3) \over \xi(-3)}  
\right]  \,,  \nonumber 
\end{eqnarray}
where $\mu$ denotes the renormalization scale, $\tilde N$= ($N$+2)/3, $\gamma$ is
Euler's constant, and $\zeta(z)$ is the Riemann zeta function. 

The three terms contributing to $F_s$ in (\ref{eq:fepertsoft}) come from one-, 
two-, and three-loop diagrams calculated in the 
effective three-dimensional theory \cite{BN95}
\begin{eqnarray}
\label{eq:lagrsoftpert}
{\cal L} &=& {\cal L}_0 + {\cal L}_{\rm int}  \,, \nonumber \\
{\cal L}_0 &=& {1\over 2} (\partial \phi_i)^2 + {1 \over 2} m^2 \phi_i^2  \,, \\
{\cal L}_{\rm int} &=& {\lambda \over 24 N} (\phi_i^2)^2 \,. 
\nonumber
\end{eqnarray}
The so-called short distance coefficients $m$ and $\lambda$ in this theory for
the soft modes are influenced by the interaction with the hard modes. Perturbatively
they are given by
\begin{equation}
  \label{eq:lambdag}
\lambda = g^2 T + o(g^4)
\end{equation}
and
\begin{eqnarray}
\lefteqn{m^2 = {1 \over 24} {\tilde N \over N} g^2(\mu) T^2 \left\{ 1 + 
{g^2 \over 16 \pi^2} \left[ - {N+8 \over 3 N} {\rm ln}{\mu \over 4\pi T}
\right. \right. }
\nonumber \\ && \left. \left. 
+ {4 \over N} {\rm ln}{\Lambda \over 4\pi T} 
- {\tilde N \over N}\gamma + 
{2 \over N} + {2 \over N} {\xi'(-1) \over \xi(-1)} \right] + o(g^4) \right\}  
  \label{eq:massg}
\end{eqnarray}
where the higher order corrections do not influence the result for the free energy
up to $O(g^5)$. Here a new parameter shows up, the separation scale $\Lambda$ 
which separates the hard from the soft modes. $\Lambda$ serves also as the 
renormalization scale in the three-dimensional theory, i.e.~it appears in the
course of renormalizing logarithmic ultraviolet divergences from loop integrals. 
Note that the short distance coefficient $m$ is not a physical quantity. If
physical quantities (like the free energy or the screening mass) are calculated 
in conventional perturbation theory the dependence of the short distance 
coefficients on $\Lambda$ exactly cancels in every order of $g$ the $\Lambda$ 
dependence coming from the renormalization of loops \cite{BN95}. Going beyond a
perturbative calculation in the following a dependence on $\Lambda$ will remain. 
This will be discussed below. 

\begin{figure}
\epsfxsize= 6.8cm
\centerline{\epsfbox{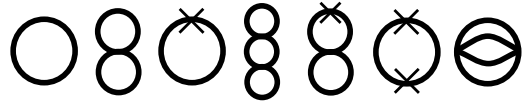}}
\vskip4mm
\caption{Feynman diagrams for the free energy contributions to be calculated in 
three-dimensional screened perturbation 
theory with massive propagators with mass $M$. The cross denotes a counter term
$-\delta m^2$.}
\label{fig:softfeyn}
\end{figure}

According to point 2 of my recipe 
I will keep the perturbative results caused by the hard modes, i.e.~the results 
for $F_h$, $m$ and $\lambda$, while according to point 3 
I will replace the three terms of $F_s$ given in (\ref{eq:fepertsoft})
by the one-, two-, and three-loop result of 
(three-dimensional) screened perturbation theory
\begin{eqnarray}
{\cal L} &=& {\cal L}_0 + {\cal L}_{\rm int}  \,, \nonumber \\
\label{eq:lagrsoftscr}
{\cal L}_0 &=& {1\over 2} (\partial \phi_i)^2 + {1 \over 2} M^2 \phi_i^2  \,, \\
{\cal L}_{\rm int} &=& - {1 \over 2} (M^2-m^2) \phi_i^2 
+ {\lambda \over 24 N} (\phi_i^2)^2 \,. 
\nonumber
\end{eqnarray}
The Feynman diagrams which have to be calculated are depicted in 
fig.~\ref{fig:softfeyn}. The cross denotes the counter term 
\begin{equation}
  \label{eq:counter}
\delta m^2 = M^2 - m^2 
\end{equation}
(times $-1$). There is an additional counter diagram not shown in 
fig.~\ref{fig:softfeyn} which serves to renormalize the mass. It cancels the
singularity arising from the basketball diagram. It also yields a finite 
contribution to the three-loop result given below. For details of the calculation
cf.~\cite{Le98}. The result for the contribution of the (resummed) soft modes to
the free energy density is
\begin{equation}
  \label{eq:fstot}
F_s = T \, (f_s^{1l} + f_s^{2l} + f_s^{3l} )
\end{equation}
with
\begin{equation}
  \label{eq:fs1l}
f_s^{1l} = -{1\over 12 \pi} N M^3  \,,
\end{equation}
\begin{equation}
  \label{eq:fs2l}
f_s^{2l} = {1\over 8 \pi} \tilde N {\lambda \over 16 \pi} M^2 
+ {1\over 8 \pi} N \delta m^2 M \,,
\end{equation}
\begin{eqnarray}
  \label{eq:fs3l}
\lefteqn{f_s^{3l} = - {1\over 8 \pi} {\tilde N^2 \over N} 
\left({\lambda \over 16 \pi} \right)^2 M 
- {1\over 8 \pi} \tilde N {\lambda \over 16 \pi} \delta m^2 }
\\ &&
- {1 \over 32 \pi} N (\delta m^2)^2 {1 \over M}
+ {1 \over 3 \pi} {\tilde N \over N} \left({\lambda \over 16 \pi} \right)^2 M 
\left( {\rm ln}{\Lambda \over 4 M} + {3 \over 2} \right)  \,.  \nonumber
\end{eqnarray}
where the appearance of a cross in a diagram is treated as being a loop, i.e.~the
third diagram of fig.~\ref{fig:softfeyn} is a two-loop contribution and the 
fifth and sixth diagram are three-loop contributions. 

The final result for the free energy density is 
\begin{equation}
  \label{eq:finfre}
F = F_h + F_s
\end{equation}
where $F_h$ is given by (\ref{eq:hardtot}-\ref{eq:Fh3l}) and $F_s$ by 
(\ref{eq:fstot}-\ref{eq:fs3l}) with the parameters given in 
(\ref{eq:lambdag},\ref{eq:massg},\ref{eq:counter}). By inspecting these equations 
we find that $F$ depends on the temperature $T$, the coupling constant $g$, the
renormalization scale (of the full four-dimensional theory) $\mu$, the separation
scale $\Lambda$, and on the so far unspecified mass $M$. In a full calculation
of the free energy density only the dependence on $T$ and $g$ would remain. In
an approximate calculation, however, there is an additional dependence on the
other three parameters and appropriate values have to be chosen for them. Since
I am mainly interested in the soft mode sector and in a proper choice for
$M$ I will determine $\mu$ and $\Lambda$ in a very simple way: I require that
perturbation theory in the hard mode sector works optimal, i.e.~I will choose
$\mu$ and $\Lambda$ such that the $g^4$ contributions to $F_h$, given in 
(\ref{eq:Fh3l}), and $m$, given in (\ref{eq:massg}), vanish. This fixes the
separation scale to
\begin{equation}
  \label{eq:sepfix}
\Lambda = 4 \pi T {\rm exp}\left( -{61\over 60}-{3\over 2}
{\zeta'(-1) \over \zeta(-1)} + {1 \over 2} {\zeta'(-3) \over \zeta(-3)} \right)
\approx 0.32 T
\end{equation}
while the renormalization scale becomes
\begin{eqnarray}
\mu &=& 4 \pi T {\rm exp}\left[ {1 \over N+8} 
\left( -(N+2)\gamma - {31\over 5} 
\right. \right. 
\nonumber \\ && \left. \left. \phantom{4 \pi T {\rm exp} m {1 \over N+8} }
- 12 {\zeta'(-1) \over \zeta(-1)} 
+ 6 {\zeta'(-3) \over \zeta(-3)} \right)  \right]  \,.
  \label{eq:renfix}
\end{eqnarray}
This yields e.g.
\begin{equation}
  \label{eq:fixmuninf}
\mu \approx 1.12 \cdot 2\pi T \quad \mbox{for}\, \, N \to \infty
\end{equation}
and
\begin{equation}
  \label{eq:fixmun1}
\mu \approx 0.57 \, T \quad \mbox{for}\, \, N = 1  \,.
\end{equation}
Admittedly, the chosen value for $\mu$ is somewhat low for the case $N = 1$ since
it is usually supposed to be somewhere around $2\pi T$ 
(cf.~\cite{BN95,BN96,DHLR97}).
However, changing $\mu$ or $\Lambda$ does not qualitatively change the results to
be presented in the following. 

Of course, the crucial question now is how to determine $M$ such that large
perturbative contributions from higher loops are resummed in the tree level mass
$M$. I will explore two different resummation procedures: \\
The first one is based on the principle of minimal sensitivity 
(see e.g.~\cite{CS98} and references therein). 
Since the full calculation for the free energy is independent of $M$ it might
be reasonable to look for an extremum of the approximate result with respect to $M$,
i.e.
\begin{equation}
  \label{eq:poms}
{\partial F \over \partial M} = 0    \qquad \mbox{(POMS)} \,.
\end{equation}
The second resummation procedure is based on the criterion of fastest apparent
convergence (see e.g.~\cite{CS98} and references therein). 
Here, an appropriate quantity is chosen, calculated
in two different orders of the approximation scheme, and it is required that the 
difference vanishes. I will use the inverse propagator evaluated at momentum
$\vec k^2 = -M^2$ and take the difference of the results for the zeroth and the 
first/second order. Since the free inverse propagator vanishes at this momentum
this simply means that the self energy is demanded to vanish at $\vec k^2 = -M^2$.
Taking the second order result for the self energy, i.e.~up to two loops, 
corresponds to the evaluation of the free energy up to three loops. This yields
the gap equation (see also \cite{CS98,Ra96,PPS97,Eb98,Pe98,Ch98} and references 
therein)
\begin{equation}
  \label{eq:feyntlg}
M^2-m^2 = 
\raisebox{-0.75em}{%
\begin{minipage}[b]{15em}
\begin{figure}
\epsfxsize= 14em
\centerline{\epsfbox{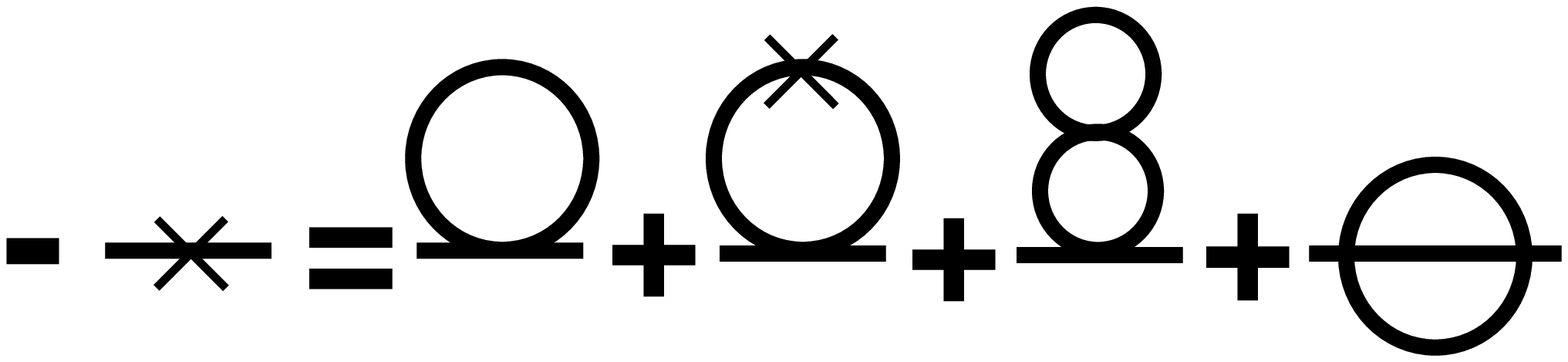}}
\end{figure}
\end{minipage}}   \vert_{\vec k^2 = - M^2}  \,.
\end{equation}
If the calculation of the free energy is restricted to maximal two loops the
corresponding one-loop gap equation results from (\ref{eq:feyntlg}) by simply 
dropping the last three diagrams. 
Again, there is an additional counter term not shown in
(\ref{eq:feyntlg}) which renormalizes the mass. It cancels the singularity arising
from the sunset diagram. For details of the calculation see \cite{Le98}. 
The result for the two-loop gap equation (TLG) is
\begin{eqnarray}
0 &=& \delta m^2 + 2 {\lambda \over 16 \pi} {\tilde N \over N} M 
- {\lambda \over 16 \pi} {\tilde N \over N} {\delta m^2 \over M}
- 2 \left({\lambda \over 16 \pi} \right)^2 {\tilde N^2 \over N^2} 
\nonumber \\ && 
+ {8 \over 3} \left({\lambda \over 16 \pi} \right)^2 {\tilde N \over N^2}
\left( {\rm ln}{\Lambda \over 8M} + {3 \over 2} \right)
\qquad \mbox{(TLG)} \,.
  \label{eq:formtlg}
\end{eqnarray}
For the one-loop gap equation (OLG) all terms on the right hand side 
except the first two have to be dropped. 

Both schemes to determine $M$ resulting in (\ref{eq:poms}) and (\ref{eq:formtlg}),
respectively, might yield more than one solution for $M$. I always take the one 
which is continuously connected to the perturbative result ($M^2 \approx m^2$). 
If the calculation of the free energy is restricted to maximal two loops the 
principle of minimal sensitivity yields the same result as the one-loop gap 
equation. In the large $N$ limit the sunset diagram (the last one in 
(\ref{eq:feyntlg})) is suppressed. There TLG, OLG, and POMS yield identical
results. This is easy to understand, if one recalls 
that for $N\to \infty$ the tadpole resummation (which is already achieved by OLG)
is sufficient to solve the exact Dyson-Schwinger equation for the propagator 
\cite{DJ74}. If the solution for $M$ is inserted in (\ref{eq:fstot}-\ref{eq:fs3l})
the three-loop contribution $f_s^{3l}$ exactly vanishes in the large $N$ limit. 
The same is true for all higher loop contributions not calculated here
(see also \cite{RS97,PKPS98}). 

For the presentation of the results one has to make a decision how to sort the
various contributions $F_h^{1l}$, $F_h^{2l}$, $F_h^{3l}$, $f_s^{1l}$, $f_s^{2l}$, 
and $f_s^{3l}$. One might be tempted to sort them in numbers of loops such that
the lowest order approximation to the free energy density would be 
$F_h^{1l}+T f_s^{1l}$, the next one $F_h^{1l}+T f_s^{1l}+F_h^{2l}+T f_s^{2l}$, 
and so on. This, however, I think is misleading in view of the schematic
picture of the perturbative contributions given in 
(\ref{eq:fepertsch},\ref{eq:fepertsoft}) and the fact that the contributions
$f_s^{1l}$, $f_s^{2l}$, and $f_s^{3l}$ only improve the soft perturbative
$g^3$, $g^4$, and $g^5$ contributions, respectively. Therefore, I choose the
following sequence of approximations: 
\begin{eqnarray}
F_0 &=& F_h^{1l} \,, \\ 
F_2 &=& F_h^{1l} + F_h^{2l} \,, \\
F_3 &=& F_h^{1l} + F_h^{2l} + T f_s^{1l}  \,, \\
F_4 &=& F_h^{1l} + F_h^{2l} + T f_s^{1l} + F_h^{3l} + T f_s^{2l}  \,, 
\label{eq:f4} \\
F_5 &=& F_h^{1l} + F_h^{2l} + T f_s^{1l} + F_h^{3l} + T f_s^{2l} + T f_s^{3l} \,.
\label{eq:f5}
\end{eqnarray}
Note that for the results shown below the renormalization scale $\mu$ is chosen 
such that $F_h^{3l}$ vanishes. Therefore the difference between 
$F_3$ and $F_4$ is solely due to $f_s^{2l}$. 

A second question which has to be clarified is which $M$ is taken for which order
of the approximation. Concerning the principle of minimal sensitivity 
for the approximation $F_5$, clearly, the mass $M$ has to be determined from
(\ref{eq:poms}) with the input $F= F_5$. With the resulting mass, $F_5$ can be 
evaluated as a function of the coupling constant $g$ (and the temperature $T$). 
One might take the same mass for the evaluation of $F_4$. However, I think it is
more appropriate here to use only $F=F_4$ as an input in (\ref{eq:poms}). This 
philosophy, however, does not work for $F_3$. Since $f_s^{1l}$ is proportional to
$M^3$, POMS would yield $M=0$, i.e.~no difference between the approximations 
$F_2$ and $F_3$. This is, of course, not very useful. Therefore I break the
rule for this special case and use for the evaluation of $F_3$ the value for $M$ 
resulting from (\ref{eq:poms}) with $F=F_4$ as input. 

\begin{figure}
\epsfxsize= 8.8cm
\centerline{\epsfbox{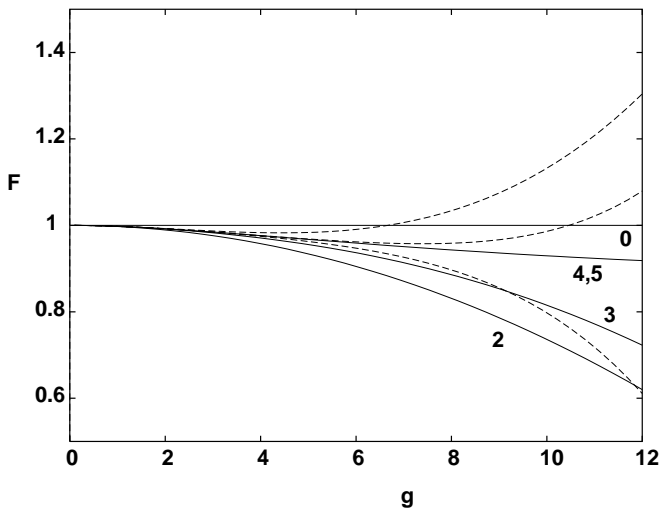}}
\vskip4mm
\caption{Various approximations $F_n$ (labeled with $n$) to the free energy density
of an $N$-component 
$\Phi^4$ theory as a function of the coupling $g$ in the large $N$ limit. See
main text for details.
The free energy density is normalized to its ideal gas value $F_0$.
The dashed lines show the perturbative approximations $O(g^3)$, 
$O(g^5)$, $O(g^4)$ (top to bottom).}
\label{fig:frenninf}
\end{figure}

For the criterion of fastest apparent convergence and its resulting one- and 
two-loop gap equations I will use the result of TLG for the evaluation of $F_5$
and of OLG for the evaluation of $F_4$. Again, it is not clear which $M$ should
be used for $F_3$. I will use the result of OLG here. 

\begin{figure}
\epsfxsize= 8.8cm
\centerline{\epsfbox{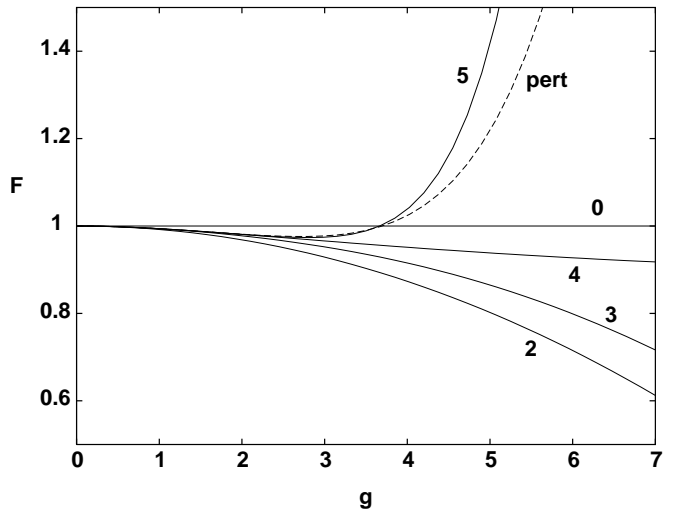}}
\vskip4mm
\caption{Various approximations $F_n$ (labeled with $n$) to the free energy density
of a one-component 
$\Phi^4$ theory as a function of the coupling $g$ using the criterion of 
fastest apparent convergence. See main text for details.
The free energy density is normalized to its ideal gas value $F_0$.
``pert'' labels the perturbative approximation $O(g^5)$.}
\label{fig:frengap}
\end{figure}

\begin{figure}
\epsfxsize= 8.8cm
\centerline{\epsfbox{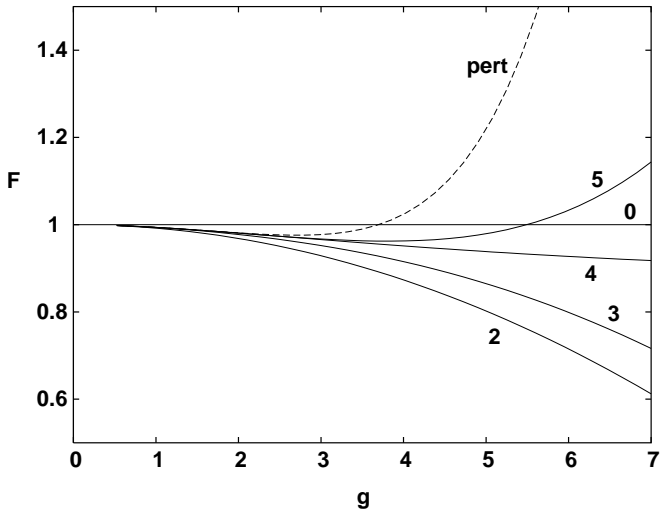}}
\vskip4mm
\caption{Same as fig.~\protect\ref{fig:frengap} but using the principle of 
minimal sensitivity.}
\label{fig:frenpoms}
\end{figure}

The results for $F_0$ - $F_5$ are shown in fig.~\ref{fig:frenninf} for the case
$N \to \infty$ and in figs.~\ref{fig:frengap} and \ref{fig:frenpoms} for $N=1$. 
To make the two cases comparable the respective maximal value of $g$ is chosen 
such that the short distance coefficient $m$ given in (\ref{eq:massg}) is 
approximately the same for both cases. For the large $N$ case I can reproduce the
results of \cite{Ka97} (cf.~fig.~\ref{fig:karsch}): Perturbation theory is improved
by the resummation scheme; even for large values of the coupling constant the
best approximation (full line in fig.~\ref{fig:karsch}, $F_5$ in 
fig.~\ref{fig:frenninf}) is quite near to the ideal gas limit (about 10\% deviation
for $g=12$); the screening mass $M$ (not shown) 
equals the perturbative one-loop result for
small coupling constants ($g^2$ part of the short distance coefficient 
(\ref{eq:massg})) and is about half of it for $g = 12$. 

Unfortunately, the picture is quite different for the case $N=1$. Comparing
the purely perturbative results (fig.~\ref{fig:pert}) with the ones obtained in
the resummation schemes (figs.~\ref{fig:frengap}, \ref{fig:frenpoms}) it turns out
that the criterion of fastest apparent convergence (fig.~\ref{fig:frengap}) does
not improve the loop expansion at all. The principle of minimal sensitivity improves
the loop expansion. However, it is striking that also in this scheme 
for larger values of the coupling constant
the approximate value for the free energy density changes from 
$\vert F \vert < \vert F_0 \vert$ to $\vert F \vert > \vert F_0 \vert$ when 
proceeding from the fourth to the fifth approximation. This seems to resemble the
oscillatory behavior of the perturbative results. Obviously, there is a window
between $g \approx 3$ and $g \approx 5$ where the latter resummation scheme (POMS)
shows a promising convergence behavior and improves the perturbative results. 
(Below $g \approx 3$ one might trust the perturbative results as well.) 
However, the fact that the two discussed resummation schemes yield
quite different results raises doubts on the reliability of both. 

Before discussing the possible lessons one could learn from the three-dimensional
version of resummed perturbation theory let me recall the additional
assumptions which enter the presented method besides the general idea
of resumming the soft modes only: First of all, I had to choose the values for
the renormalization scale $\mu$ and the separation scale $\Lambda$. I have used
a very simple recipe here to fix these quantities by demanding that the 
$g^4$ contributions to the hard mode quantities $F_h$ and $m$ have to vanish
(cf.~(\ref{eq:sepfix}-\ref{eq:fixmun1}) and discussion there). It is worth noting
that this approach is
in accordance with the criterion of fastest apparent convergence which I have used 
later on. In principle,
if perturbation theory works these $g^4$ contributions would be small but
in general would not completely vanish. Therefore, forcing them to vanish
by choosing the scales accordingly might cause other contributions to become 
unnaturally large. On the other hand, this simple choice allows to discuss
in the cleanest way
the convergence behavior of the soft mode sector and its influence on the total
result for the free energy. If e.g.~$\mu$ is chosen in another way then
the influence of $f_s^{2l}$ and $f_s^{3l}$ in (\ref{eq:f4},\ref{eq:f5}) would
intertwine with the influence of $F_h^{3l}$. Nonetheless, it is important to
check whether at least the qualitative results are robust against changes in
$\mu$ and $\Lambda$. Therefore, I will now explore other choices for the scale 
parameters. I will restrict myself to the most interesting case $N=1$.

\begin{figure}
\epsfxsize= 8.8cm
\centerline{\epsfbox{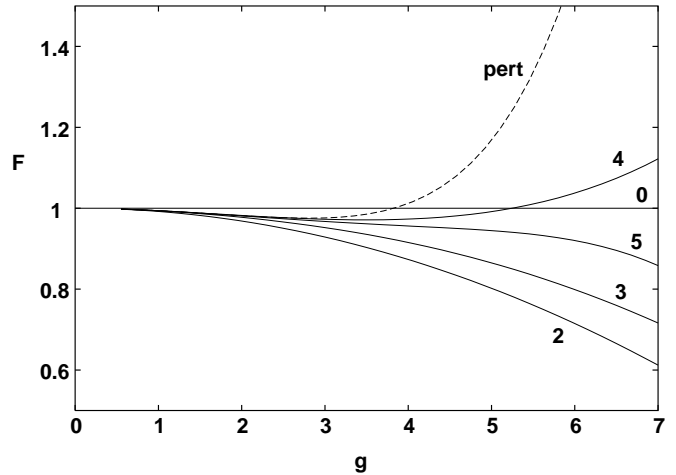}}
\vskip4mm
\caption{Same as fig.~\protect\ref{fig:frengap} but using the principle of 
minimal sensitivity and $\mu = T$.}
\label{fig:mut}
\end{figure}

Concerning the renormalization scale a reasonable choice
might be $\mu = 2 \pi T$ since this is the typical energy scale of the hard modes
\cite{BN95,BN96,DHLR97}. For this case the contribution of the
hard modes to the free energy (\ref{eq:hardtot}) becomes (for $N=1$)
\begin{equation}
  \label{eq:mu2pit}
F_h \approx 
F_h^{1l} \left[ 1 - 0.0079 \, g^2 \left( 1 - 0.046 \, g^2 \right) \right]  
\quad \mbox{for} \quad \mu = 2 \pi T \,.
\end{equation}
One finds that the $g^4$ contribution overwhelms the $g^2$ contribution already
for $g \approx 4.7$. In view of the large coupling constants 
I have dealt with in this
work (up to $g = 7$ for $N=1$) this choice of $\mu$ is inappropriate since in this
case the naive perturbation theory which I have used in the hard mode sector breaks
down. For the more naive choice $\mu = T$ one gets a $g^4$ coefficient which 
is somewhat smaller:
\begin{equation}
  \label{eq:mut}
F_h \approx 
F_h^{1l} \left[ 1 - 0.0079 \, g^2 \left( 1 - 0.011 \, g^2 \right) \right]  
\quad \mbox{for} \quad \mu = T \,.
\end{equation}
For this choice of $\mu$ (and $\Lambda$ as given in (\ref{eq:sepfix})) 
the free energy contributions are plotted in fig.~\ref{fig:mut}
using the principle of minimal sensitivity which for the original choice of 
$\mu$ was
the better one as compared to the criterion of apparent convergence. Obviously, also
for this choice of the renormalization scale POMS improves the convergence as 
compared
to the perturbative loop expansion. However, one finds qualitatively the same 
feature as
in fig.~\ref{fig:frenpoms}, namely the oscillatory behavior around the ideal gas 
limit
for coupling constants larger than $g \approx 5$. Note, however, that the fourth
and fifth approximation basically have now exchanged their places as compared to 
fig.~\ref{fig:frenpoms}. 

\begin{figure}
\epsfxsize= 8.8cm
\centerline{\epsfbox{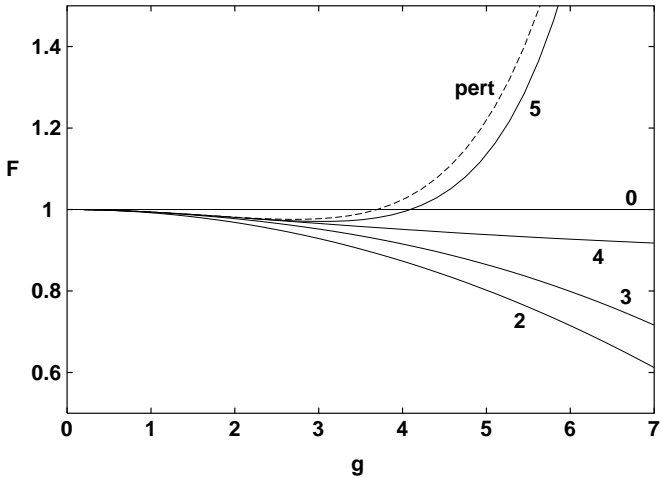}}
\vskip4mm
\caption{Same as fig.~\protect\ref{fig:frengap} but using the principle of 
minimal sensitivity and $\Lambda = \sqrt{2 \pi T \, m_{1l} } $.}
\label{fig:lageme}
\end{figure}

Concerning 
the separation scale it is reasonable to put it somewhere between the typical 
energy scales of the hard and the soft modes. 
While the former is $2 \pi T$ as discussed above the latter is characterized by
$m$ or roughly the lowest perturbative ($g^2$) contribution to $m$ 
(as long as perturbation theory holds for the hard mode sector):
\begin{equation}
  \label{eq:m1l}
m_{1l}^2 := {1 \over 24} {\tilde N \over N} g^2(\mu) T^2  \,.
\end{equation}
Thus a reasonable choice is
\begin{equation}
  \label{eq:lambdageommean}
\Lambda = \sqrt{2 \pi T \, m_{1l} } \sim \sqrt{g}   \,.
\end{equation}
Fig.~\ref{fig:lageme} shows the free energy contributions for this choice of the 
separation
scale (and $\mu$ as given in (\ref{eq:fixmun1})) using again the principle of 
minimal 
sensitivity. Comparing the values for $F_5$ from fig.~\ref{fig:frenpoms} and 
fig.~\ref{fig:lageme} shows that the latter choice for the separation scale 
(\ref{eq:lambdageommean}) makes the convergence behavior worse. 
In addition, it turns out that in this case
the full POMS condition, i.e.~using $F=F_5$ in (\ref{eq:poms}), 
does not have a solution
for $M$ for coupling constants smaller than $g \approx 2$.
Therefore, in fig.~\ref{fig:lageme} $F_5$ is plotted only for $g > 2$ 
(hardly visible).
All of this is
qualitatively in line with the general finding that the presented resummation scheme
does not improve the loop expansion for the case $N =1$. 

Besides the specific choices for the renormalization and separation scales I 
have tacitly assumed that there is only one counter term (\ref{eq:counter}) in
(\ref{eq:lagrsoftscr}). Alternatively one could introduce an additional counter
term for each additional loop order. This approach will be discussed in more
detail in \cite{Le98}. Qualitatively this modified method yields the same results
as presented here. Especially no further improvement concerning the poor
convergence behavior for $N=1$ is achieved by this modification.

\begin{figure}
\epsfxsize= 8.8cm
\centerline{\epsfbox{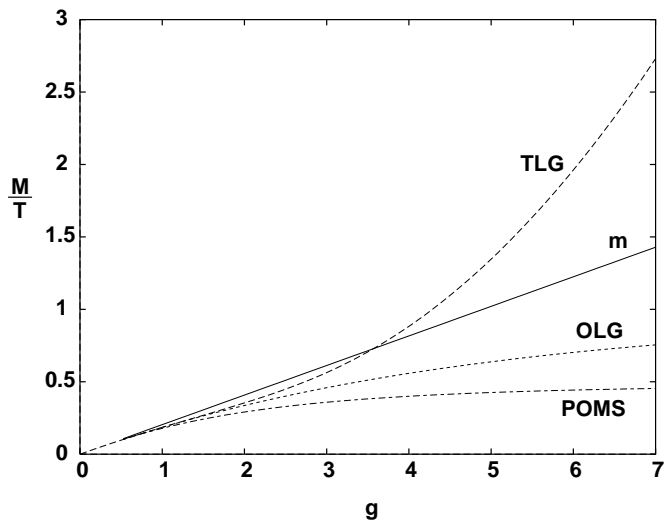}}
\vskip4mm
\caption{The resummed mass $M$ as obtained from POMS, TLG, and OLG and the short
distance coefficient $m$ as a function of the coupling $g$ for a one-component 
$\Phi^4$ theory. 
See main text for details.}
\label{fig:masses}
\end{figure}

To summarize, I have found that a scheme which combines conventional perturbation 
theory for the hard modes with screened perturbation theory for the soft modes 
significantly improves conventional perturbation theory for an $N$-component 
$\Phi^4$ theory in the large $N$ limit. Contrary, this scheme does
not work equally well for finite $N$. There are several possibilities to explain
this shortcoming: 

1. The presented scheme is based on the assumption that the hard and the soft
modes can be separated and that the screening effect can be neglected for the 
hard modes. The necessary condition for this assumption to hold is that in the
Matsubara propagator $1/(\vec k^2 + (2n\pi T)^2 + M^2)$ the mass term can be 
neglected compared to the temperature term for $n \neq 0$, i.e. 
\begin{equation}
  \label{eq:masstemp}
M^2 \ll (2 \pi T)^2      \,.
\end{equation}
If the screening mass does not fulfill this condition the presented scheme
would not be appropriate. In turn, if screened perturbation theory works in the
four-dimensional theory and yields a screening mass which fulfills 
(\ref{eq:masstemp}) then also my three-dimensional version of screened perturbation
theory should work. Of course, I cannot say what the true value for the 
screening mass is. However, I can at least check whether the result for $M$ 
obtained in my approximation scheme is consistent with the underlying assumption 
(\ref{eq:masstemp}). Fig.~\ref{fig:masses} shows the results for $M$ for
the different resummation schemes. The line labeled with ``POMS'' is obtained
from (\ref{eq:poms}) using the ``best'' result $F = F_5$ from (\ref{eq:f5}). The
``TLG'' and ``OLG'' lines result from (\ref{eq:formtlg}) with the appropriate
terms dropped in the latter case. I think it is fair to say that the POMS and
OLG results fulfill (\ref{eq:masstemp}) in the plotted regime for the coupling 
constant. For the TLG result relation (\ref{eq:masstemp}) is not very well satisfied
for the largest plotted values of the coupling constant. However, for $g \approx 5$
there is a factor of about 40 between $M^2$ and $(2\pi T)^2$ so that 
(\ref{eq:masstemp}) is roughly fulfilled. Fig.~\ref{fig:frengap} shows that already
here the convergence of the resummation scheme is worse than the conventional
perturbative scheme. Thus, the failure of the resummation scheme utilizing the
criterion of fastest apparent convergence cannot be traced back to its inconsistency
with the underlying assumption (\ref{eq:masstemp}). 

2. Another reason for the finding that the presented resummation scheme does not
work equally well for $N \to \infty$ and $N=1$ may be that only a momentum
independent mass is resummed. It might appear that the resummation of a more
general, i.e.~momentum dependent
self energy (as e.g.~advocated in \cite{RS97}) would be more appropriate. 
For the large $N$ case this would not change anything since the sunrise diagram
and all higher momentum dependent diagrams are suppressed by powers of $1/N$. 
As already pointed out before, the resummation of tadpoles is the only thing
which has to be done in the large $N$ limit. If the possible momentum dependence
is a necessary ingredient for a reliable resummation scheme then this would 
naturally explain why the scheme presented here works for $N \to \infty$ but fails
for $N =1$. I will elaborate on this question in \cite{Le98}. 

3. Besides its momentum dependence the sunrise diagram (and the corresponding
basketball diagram contributing to the free energy) gives rise to a logarithmic
term in the mass $M$. 
One may argue that this logarithmic
term via an inappropriately large contribution ${\rm ln}(\#\Lambda/M)$ 
shifts the mass to a wrong place and causes a breakdown of the 
three-dimensional screened perturbation theory. Since this logarithmic term is
caused by the renormalization of the theory it might be possible to handle
that problem by renormalization group equations (cf.~e.g.~\cite{BN95}). 
Of course, this is a pure speculation so far and deserves further studies. If it
is true it would also explain why the resummation works in the large $N$ limit.
There, these logarithmic contributions are suppressed as one can easily check
by inspecting (\ref{eq:fs3l}) and (\ref{eq:formtlg}).

4. The most negative explanation for the shortcomings of the presented
resummation scheme would be that the concept of screened perturbation theory
does not work at all (or only for very special cases like the large $N$ limit)
or more generally that concerning the determination of the free energy 
there is so far no method which reorganizes the loop expansion in a
useful way, at least for coupling constants which are not tiny. 
This would imply that we have to rely on genuine non-perturbative
approaches like lattice calculations. Indeed, the free energy density is a quantity
which can be calculated on the lattice (see e.g.~\cite{Dehwa2,LaerQM}). 
However, for many other,
e.g.~non-static quantities there are fundamental obstacles which so far prevent 
their evaluation on the
lattice. If (improved) perturbation theory fails to calculate
the free energy density one may also doubt the results for other
quantities obtained in a loop expansion in high temperature quantum 
field theory. 

I conclude that the question how to reliably calculate the free energy density in a 
modified loop expansion is still open.

\acknowledgements

I thank Eric Braaten and Agustin Nieto for introducing to me their effective field
theory method. I also thank the organizers of the 
``5th International Workshop on Thermal Field Theories and Their Applications''
for providing the opportunity for very interesting discussions.


\begin{references}
\bibitem{BN95} E.~Braaten and A.~Nieto, Phys.~Rev.~{\bf D51}, 6990 (1995).
\bibitem{Ka97} F.~Karsch, A.~Patk\'os, and P.~Petreczky, Phys.~Lett. {\bf B401}, 
69 (1997). 
\bibitem{CP94} C.~Coriano and R.R.~Parwani, Phys.~Rev.~Lett.~{\bf 73}, 2398 (1994).
\bibitem{Pa94} R.R.~Parwani, Phys.~Lett.~{\bf B334}, 420 (1994), 
Erratum-ibid.~{\bf B342}, 454 (1995).
\bibitem{PC94} R.R.~Parwani and C.~Coriano, Nucl.~Phys.~{\bf B434}, 56 (1995).
\bibitem{AZ94} P.~Arnold and C.~Zhai, Phys.~Rev.~{\bf D50}, 7603 (1994).
\bibitem{AZ95} P.~Arnold and C.~Zhai, Phys.~Rev.~{\bf D51}, 1906 (1995).
\bibitem{ZK95} C.~Zhai and B.~Kastening, Phys.~Rev.~{\bf D52}, 7232 (1995).
\bibitem{BN96} E.~Braaten and A.~Nieto, Phys.~Rev.~{\bf D53}, 3421 (1996).
\bibitem{AHLW97} M.~Achhammer, U.~Heinz, S.~Leupold, and U.A.~Wiedemann, 
Ann.~Phys.~(NY) {\bf 261}, 1 (1997).
\bibitem{FST92} J.~Frenkel, A.V.~Saa, and J.C.~Taylor, 
Phys.~Rev.~{\bf D46}, 3670 (1992).
\bibitem{PS95} R.~Parwani and H.~Singh, Phys.~Rev.~{\bf D51}, 4518 (1995).
\bibitem{KLPR97} F.~Karsch, M.~L\"utgemeier, A.~Patk\'os, and J.~Rank, 
Phys.~Lett.~{\bf B390}, 275 (1997).
\bibitem{Kast97} B.~Kastening, Phys.~Rev.~{\bf D56}, 8107 (1997).
\bibitem{Ha97} T.~Hatsuda, Phys.~Rev.~{\bf D56}, 8111 (1997).
\bibitem{DHLR96} I.T.~Drummond, R.R.~Horgan, P.V.~Landshoff, and A.~Rebhan,
Phys.~Lett.~{\bf B398}, 326 (1997).
\bibitem{RS97} J.~Reinbach and H.~Schulz, Phys.~Lett.~{\bf B404}, 291 (1997).
\bibitem{DHLR97} I.T.~Drummond, R.R.~Horgan, P.V.~Landshoff, and A.~Rebhan,
Nucl.~Phys.~{\bf B524}, 579, (1998).
\bibitem{PKPS98} A.~Peshier, B.~K\"ampfer, O.P.~Pavlenko, and G.~Soff,
Europhys.~Lett.~{\bf 43}, 381 (1998).
\bibitem{VB98} B.~Vanderheyden and G.~Baym, hep-ph/9803300.
\bibitem{BLNR98} D.~B\"odeker, P.V.~Landshoff, O.~Nachtmann, and A.~Rebhan,
hep-ph/9806514.
\bibitem{Re98} A.~Rebhan, these proceedings, hep-ph/9808480. 
\bibitem{Sc98} H.~Schulz, these proceedings, hep-ph/9808339. 
\bibitem{Creutz} M.~Creutz, {\sl Quarks, Gluons, and Lattices}
(Cambridge University Press, Cambridge, 1983).
\bibitem{Ka89} F.~Karsch, Nucl.~Phys.~{\bf B9} [Proc.~Suppl.] 357 (1989).
\bibitem{GS93} V.~Goloviznin and H.~Satz, Z.~Phys.~{\bf C57}, 671 (1993).
\bibitem{PK96} A.~Peshier, B.~K\"ampfer, O.P.~Pavlenko, and G.~Soff, 
Phys.~Rev.~{\bf D54}, 2399 (1996).
\bibitem{LH97} P.~Levai and U.~Heinz, Phys.~Rev.~{\bf C57}, 1879 (1998).
\bibitem{FKRS94} K.~Farakos, K.~Kajantie, K.~Rummukainen, and M.~Shaposhnikov,
Nucl.~Phys.~{\bf B425}, 67 (1994).
\bibitem{Le98} S.~Leupold, in preparation.
\bibitem{CS98} S.~Chiku and T.~Hatsuda, Phys.~Rev.~{\bf D58}, 076001, (1998).
\bibitem{Ra96} R.O.~Ramos, Braz.~J.~Phys.~{\bf 26}, 684 (1996), hep-ph/9607416.
\bibitem{PPS97} A.~Patk\'os, P.~Petreczky, and Zs.~Sz\'ep, Eur.~Phys.~J.~{\bf C5}, 
337 (1998).
\bibitem{Eb98} F.~Eberlein, Phys.~Lett.~{\bf B439}, 130 (1998).
\bibitem{Pe98} P.~Petreczky, these proceedings, hep-ph/9809414. 
\bibitem{Ch98} S.~Chiku, these proceedings, hep-ph/9809215.
\bibitem{DJ74} L.~Dolan and R.~Jackiw, Phys.~Rev.~{\bf D9}, 3320 (1974). 
\bibitem{Dehwa2} C.~DeTar, in {\sl Quark-Gluon Plasma 2}, p.~1, edited by R.C.~Hwa 
(World Scientific, Singapore, 1995). 
\bibitem{LaerQM} E.~Laermann, Nucl.~Phys.~{\bf A610}, 1c (1996). 
\end{references}
\end{document}